# Nonvolatile single-ion memory with picosecond switching

Hengxiao Cheng[1], Xudong Zhu[2,3], Zijia Su[1,4], Zhongbin Dai[1], Jie Yu[1], Zhi Yan[5], Xujin Zhang[5], Renfa Zhou[1], Juan Wang[1], Yuanyuan Shi[1], Zhongguang Xu[1]*, Lixin He[2,3]*, Chengjie Zuo[1]*

[1]School of Integrated Circuits, University of Science and Technology of China, Hefei, China

[2]Institute of Artificial Intelligence, Hefei Comprehensive National Science Center, Hefei, China

[3]CAS Key Laboratory of Quantum Information, University of Science and Technology of China, Hefei, China

[4]School of Integrated Circuit, Hefei University of Technology, Hefei, China

[5]School of Chemistry and Materials Science & Key Laboratory of Magnetic Molecules and Magnetic Information Materials of Ministry of Education, Shanxi Normal University, Taiyuan, China

*e-mail: xuxu@ustc.edu.cn, helx@ustc.edu.cn, czuo@ustc.edu.cn

## Abstract

The rapid development of artificial intelligence (AI), Internet of Things (IoT), and edge computing applications has posed severe challenges to conventional memory technologies in terms of density, speed, and energy consumption. Herein, a single-ion transport mechanism is proposed to achieve picosecond (ps) switching capability. For monolayer hexagonal boron nitride (h-BN) with single-atom vacancy defects, first-principles calculations reveal that single-ion penetration across the BN plane dominates the resistive switching. The trapping and release of a single ion correspond to different states of the memory device for one bit of information. Experimentally fabricated single-ion memory exhibits nonvolatile resistive switching with ultra-fast switching speed of 20 ps and ultra-low energy consumption of 310 aJ/bit. This high performance is attributed to the extremely short distance for the single ion to travel through. Such devices pave the way for the realization of high-performance nonvolatile memory with ultra-fast speed, ultra-low energy consumption, and high storage density, that is called the "Unified Memory" long desired by the whole industry.

# Main

Across ages, the quest to store information in ever-diminishing units has been a constant pursuit[1]. Reducing the memory cell size can not only improve information storage density, but also deliver faster speed and lower energy consumption[2]. The mainstream commercial memories, including Flash[3], dynamic random access memory (DRAM)[4], and static random access memory (SRAM)[5], can hardly simultaneously achieve high speed, nonvolatility, and high density[6]. Although three-dimensional (3D) integrated flash[7], phase-change memory (PCM)[8], and magnetoresistive random access memory (MRAM)[9] have emerged successively, their further development is fundamentally limited by factors such as device structure and operating conditions. With exponentially growing demands of the global datasphere and artificial intelligence (AI) computing, there is an urgent need to explore novel memory technologies with simple structure, facile operation, compatibility with the complementary metal oxide semiconductor (CMOS) processes, and especially fast speed, so as to break the "Memory Wall"[6] for higher data bandwidth and lower energy consumption.

As a promising candidate technology, resistive random access memory (RRAM) emulates biological synapses, enabling low-power, high-speed, and nonvolatile storage[10]. With the rapid rise of two-dimensional (2D) materials, novel types of RRAMs have been widely reported[11-13] to have low operating voltage, fast switching speed, large ON/OFF ratio, and compatibility with CMOS. Compared to other 2D materials, hexagonal boron nitride (h-BN) is considered to be one of the most suitable candidates for RRAM implementation[14,15] due to its large bandgap (~6 eV), high heat thermal conductivity, and excellent stability.

As Dr. Richard P. Feynman proposed in his famous talk "There's Plenty of Room at the Bottom (1959)"[1], a bit of information could be stored in a cube of atoms 5 times 5 times 5 – that is 125 atoms. We take a step forward in this paper and represent a bit of information by manipulating a single ion. Monolayer of chemical vapor deposition (CVD) grown h-BN with single-atom vacancy defects[16] was used in the following demonstration. The resistive switching mechanism of monolayer h-BN based memory was firstly investigated by doing first-principles calculations. As is discovered, the single-ion transport (SIT) occurring at a boron vacancy defect to form a dual-ion bridge conductive path (CP) plays a significant role in the switching process. Leveraging this mechanism, we fabricated nonvolatile bipolar resistive switching (BRS) devices with CVD-grown monolayer h-BN in a metal-insulator-metal (MIM) sandwich structure. The devices exhibit large ON/OFF ratio >$10^5$, retention >10,000 s and endurance ~400 cycles when multiple conductive paths operate as a whole. In the case of activating only a single conductive path by confining programing energy, due to the sub-nanometer migration path in the single-ion transport process and sub-quantum conductance, the devices achieve ultra-fast switching (~20 ps) and ultra-low energy consumption (~310 aJ/bit). This newly discovered SIT mechanism – trapping a single ion as a single-bit nonvolatile memory device approaching the ultimate fast-speed and low-energy switching limit – provides a new technology platform for implementing the "Unified Memory or Universal Memory"[6] with nonvolatility, fast speed, and high density.

**The single-ion transport (SIT) switching**

The formation and rupture of conductive filaments (CFs) are universally recognized as the key to resistive switching. However, when the switching layer is scaled to a single atomic layer, the conductive path (CP) formed at a single atom vacancy reduces to a simple quasi-dot. The switching mechanism therefore demands further investigation. Among candidate materials, monolayer h-BN grown by CVD intrinsically contains single atom vacancies[16], providing an ideal platform to probe this ultimate regime. It is also amenable to scalable industrial production[17]. The mechanism of the resistive switching process in h-BN is elucidated by performing first-principles calculations.

With titanium (Ti) metal ion as the exemplar, Fig. 1a illustrates the schematic of the single-ion transport across a single-atom vacancy defect. The negligible energy barriers of diffusion on a defect-free h-BN surface (~0.3 eV, Supplementary Fig. 1 and Supplementary Video 1) and the energy minimum located at boron vacancy ($V_B$) defect (Supplementary Fig. 2) demonstrate that the Ti ions are spontaneously trapped at defect sites. It is noteworthy that there are two relaxed sites with energy minimum located on either side of the defect. To migrate from one relaxed site through the BN plane to another relaxed site, Ti ion needs to push away the surrounding nitrogen atoms around $V_B$ (Fig. 1a, Supplementary Fig. 2b, and Supplementary Video 2). An energy barrier exists during this process, thus necessitating an external driving electric field.

Moreover, calculations regarding the various defect types were also performed. Different from the $V_B$ single-vacancy defect, the energy barrier for $Ti^{3+}$ entering a nitrogen vacancy ($V_N$) defect (~0.7 eV, Supplementary Fig. 3) is higher than the barrier for leaving (~0.4 eV). It can be attributed to the positive charge of $V_N$. Since Ti ion needs to repel surrounding boron atoms when moving through the $V_N$ defect, the migration energy is extremely high (~5.0 eV, Supplementary Fig. 4). Therefore, Ti ion migration at $V_B$ defect, rather than $V_N$ defect, plays a dominant role in the switching process.

Fig. 1b and Supplementary Video 7 display the time-resolved migration process dynamics in monolayer h-BN, calculated from first-principles. The single-atomic thickness of h-BN maximizes the electric field generated under bias. According to the calculations, within 8 picosecond (ps), one Ti ion is able to penetrate the $V_B$ defect and settle at the bottom relaxed site (defined as the Trap site), with another ion occupying the top relaxed site (defined as the Base site), as shown in the insets of Fig.1b. Based on the formula of high-field ionic drift[18], the ion migration average velocity is given by:

$$v \approx fa\exp\left(-\frac{E_b}{kT}\right)\sinh\left(\frac{qEa}{2kT}\right) \quad (1)$$

where $a$, $E_b$, $E$, and $T$ represent the distance between the two relaxed sites, energy barrier, electric field, and absolute temperature, while $f$, $q$, and $k$ correspond to the frequency of penetrate attempts, charge of an electron, and Boltzmann's constant, respectively. Therefore, the ultra-fast switching stems from the ultimately short migration distance and the maximized electric field.

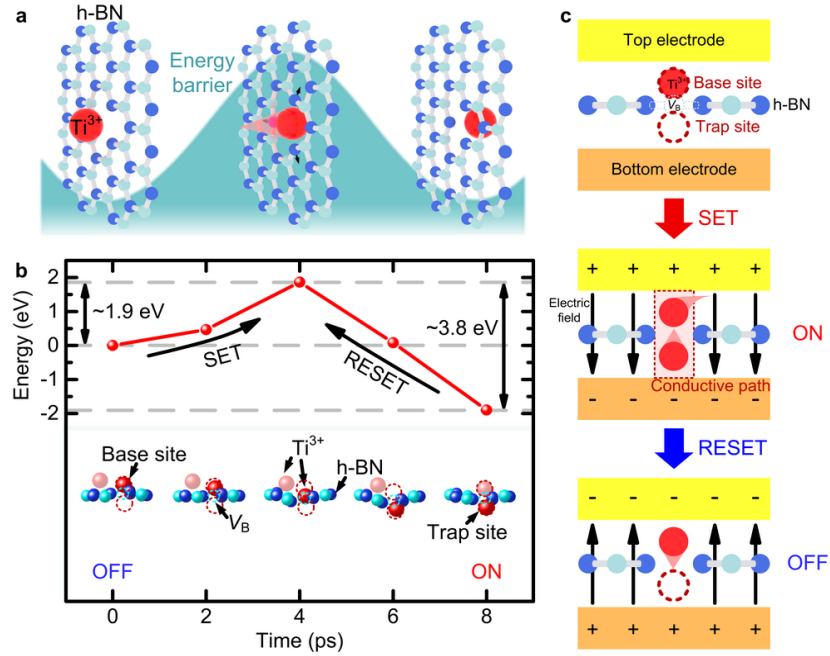

**Fig. 1: The single-ion transport switching in monolayer h-BN. a**, The schematic of the single-ion transport switching. The blue background represents the energy variation. The defect in monolayer h-BN is flanked by two minimum energy sites (defined as relaxed site). Ti ion is initially trapped at one relaxed site. To transport across BN plane and reach the relaxed site on the opposite side, it is necessary to overcome the energy barrier caused by repelling surrounding atoms. **b**, The first-principles calculations of complete switching process. One Ti ion penetrates across the BN plane and the vacated Base site is replenished by another ion as shown in the insets. The migration process completed at the 8th picosecond. The RESET energy barrier is larger than that in SET process. It demonstrates that more energy is required to RESET. The calculated dynamic distribution is shown in Supplementary Videos 5,7. **c**, The diagram of the resistive switching in single-ion memory. The trapping and release of single Ti ion in the Trap site correspond to ON and OFF states, which can be utilized as "1" and "0" for one bit of information.

As shown in Fig. 1c, when a single Ti ion is trapped at the Trap site and another Ti ion refilling the Base site, the two ions (a dual-ion bridge) form a conductive path (CP) that goes through the monolayer h-BN and the memristive device is turned ON with higher conductance. Otherwise, the device remains in the OFF state with lower conductance. The different states can be utilized as "1" and "0" in memory. One bit of information is stored depending on the presence or absence of a single Ti ion at the Trap site. Considering the area of memory cell (3 hexagonal rings) is about 0.16 nm$^2$, the theoretical storage density limit can be up to 6 bit/nm$^2$. The penetration barrier is about 1 ~ 1.9 eV, which varies depending on the distribution of the Ti ions (Fig. 1b and Supplementary Fig. 5). The larger reverse barrier for the ion to transport from the Trap site back to the Base site implies that the RESET process demands more energy and is assisted by Joule heating[11].

## Characterizations of CVD monolayer h-BN based resistive switches

Ions penetrating $V_B$ defects in monolayer h-BN face a low energy barrier and an ultra-short distance. It therefore provides a theoretical basis for achieving ultra-low energy consumption and ultra-fast switching. To verify the single-ion transport (SIT) mechanism, CVD-grown monolayer h-BN based resistive switches were fabricated.

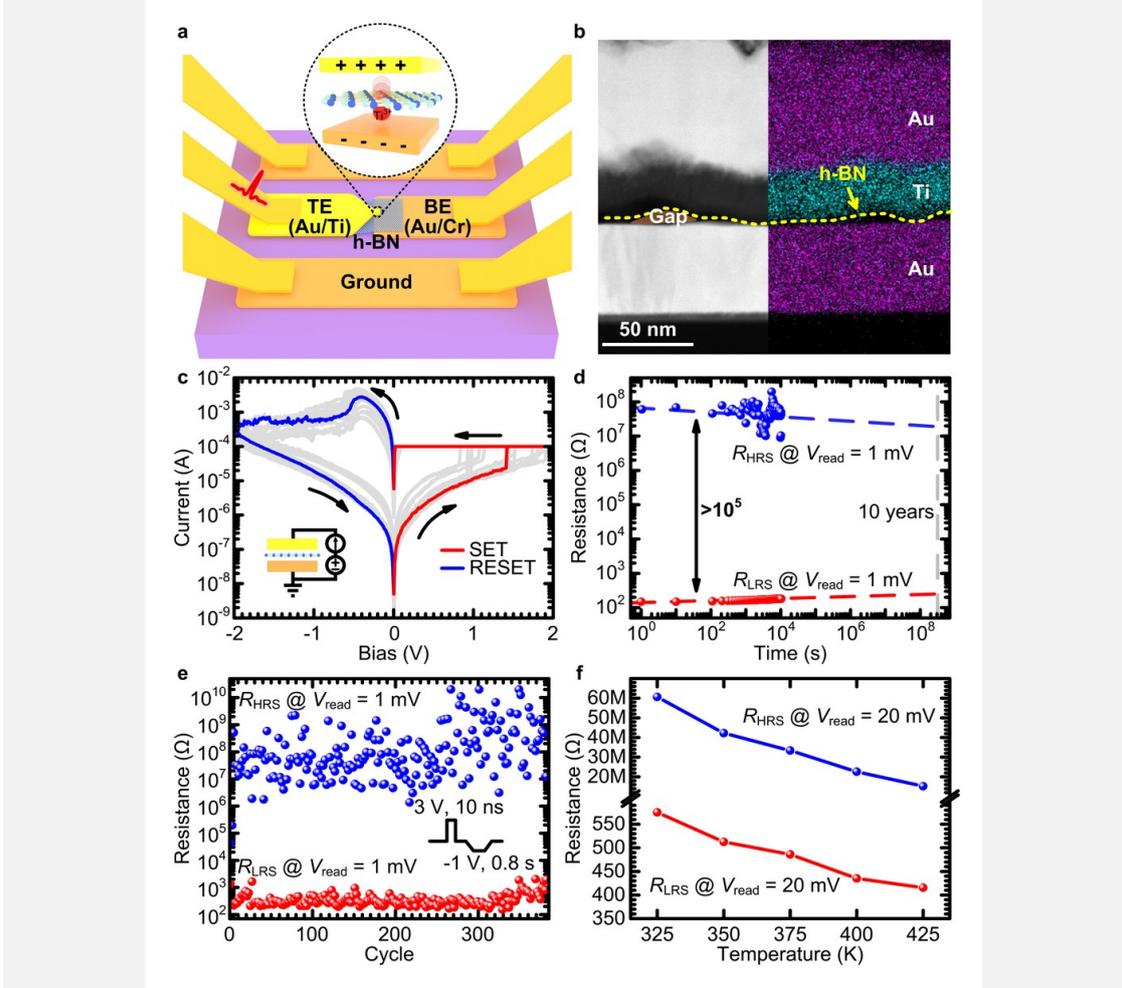

**Fig. 2: Structure and characterizations of CVD monolayer h-BN based devices. a**, The schematic of h-BN based switches. A ground-signal-ground structure was employed to achieve 50 Ω impedance matching. **b**, Cross-sectional scanning transmission electron microscopy (STEM) image (left) and energy dispersive spectroscopy (EDS) analysis (right) of the Au/Ti/h-BN/Au stack. The transfer-induced gap exists between the BE and h-BN. **c**, The repeated *I-V* curves of CVD monolayer h-BN based BRS, measured by the circuit as shown in the inset. **d**, The retention result with $V_{read}$ = 1 mV exhibits good nonvolatility that shows no deterioration beyond 10,000 s. It can be extended to approximately 10 years or longer. **e**, The endurance result with $V_{read}$ = 1 mV shows normal switching for about 400 cycles under applied pulses. **f**, The dependence of resistance on temperatures, reading from Supplementary Fig. 9. The negative temperature coefficients observed in both HRS and LRS indicate their semiconductor characteristics, as their resistances decrease with increasing temperature.

Fig. 2a shows the schematic of the h-BN based switches. Ground-signal-ground (GSG) probing pads were adopted to realize high-frequency 50 Ω impedance matching, enabling pulse-test characterization of ultra-fast switching speed. The scanning electron microscopy (SEM) images are shown in Supplementary Fig. 6a. The CVD-grown monolayer h-BN was wet-transferred onto a sapphire substrate with pre-deposited bottom electrode (BE, 5 nm Cr/50 nm Au). The excess h-BN is etched by reactive ion etching (RIE) after lithographic patterning. After that, the top electrode (TE, 20 nm Ti/60 nm Au) was deposited by electron beam evaporation after lithograph (see Methods and Supplementary Fig. 7 for fabrication details).

Fig. 2b displays cross-sectional scanning transmission electron microscopy (STEM) image and energy dispersive spectroscopy (EDS) analysis of the Au/Ti/h-BN/Au stack. The gap between the BE and h-BN can be attributed to the fact that the h-BN established van der Waals (vdW) contact with the BE during the transfer process, while the TE was deposited on the h-BN with no gap. Due to the existence of gap between the BE and h-BN, Ti ions in the TE migrate more easily into h-BN defects than Au in the BE. Supplementary Fig. 6b shows the Raman spectrum, and the peak at ~1370 cm$^{-1}$ corresponds to the $E_{2g}$ phonon vibration mode of h-BN. The AFM morphology image at the edge of monolayer h-BN (Supplementary Fig. 6c) demonstrates the monolayer thickness of ~0.4 nm, which is consistent with previous reports[19].

**Bipolar resistive switching in Au/Ti/h-BN/Au stack**

Prior to exploration of the proposed single-ion memory, the fabricated devices were measured to demonstrate the normal memristive characteristics with multiple conductive paths (dual-ion bridges) operating as a whole. The electrical properties of the CVD-grown monolayer h-BN based devices were characterized by applying bias to the TE and grounding the BE, as shown in the inset of Fig. 2c. The devices with Au/Ti/monolayer h-BN/Au stack exhibit bipolar resistive switching (BRS) phenomenon (Figs. 2c-e), with large ON/OFF ratio >10$^5$, long retention >10,000 s without noticeable degradation, and endurance ~400 cycles with wide-pulse sequence programming.

At the beginning, the BRS is in a high resistance state (HRS). When the positive bias is swept forward to a threshold, Ti ions migrate through the defects to the Trap sites, which would form conductive dual-ion bridges between the TE and BE. The BRS switches to a low resistance state (LRS), and the current increases sharply (SET process). During this process, the current compliance ($I_{cc}$) was set to 100 $\mu$A to protect the device from breakdown. The minor discrepancy of switching voltage may be due to Joule heating and defect states. This is confirmed by the first-principles calculations shown in Fig. 1b and Supplementary Fig. 5. Differences in the initial distributions of Ti ions give rise to variations in the calculated energy barriers. The barrier height corresponds to the required switching voltages. The experimental $V_{SET}$ is consistent with first-principles calculation results, confirming that Ti ion migration across the $V_B$ defect dominates the resistive switching process.

Until a sufficient negative bias is applied to the BRS, the conductive bridges are then ruptured under the combined effect of electric field driving and Joule heating[11].

The Ti ions at the Trap sites depart, leaving the sites vacant. As a result, the BRS returns to HRS (RESET process). Ions migrate individually during this process, so it is promising to achieve multi-level storage by controlling the sweep waveform (Supplementary Fig. 8)[20].

Moreover, as visualized in Fig. 2f and Supplementary Fig. 9, the resistance decreases with rising temperature in the LRS region. The semiconductor-like behavior can be attributed to Ti ions rather than neutral metallic atoms[21,22]. Meanwhile, the reduced HRS resistance illustrates the semiconducting property with a lower barrier in high temperature range[11,20].

The cumulative probability plot in Supplementary Figs. 10,11 reveal acceptable cycle-to-cycle and device-to-device variations. The dominating conduction mechanism transitions from ohmic conduction to Schottky emission[23] at 80 mV, as shown in Supplementary Fig. 12. In addition to the Au/Ti/h-BN/Au stack, devices without h-BN (i.e., Au/Ti/Au, Supplementary Fig. 13a) and with CVD monolayer graphene (i.e., Au/Ti/graphene/Au, Supplementary Fig. 13b) were also fabricated for comparison and they didn't exhibit notable memristive *I-V* behavior, which verifies that the BRS behavior stems from h-BN rather than metal oxides or impurities introduced during the transfer process.

**Sub-quantum conductance in a single conductive dual-ion bridge**

Leveraging the defect-specific variation of $V_{SET}$ arising from diverse distributions of Ti ions, different conductance levels $R_{LRS}^{-1}$ can be attained by setting a small $I_{cc}$ and reducing the sweep voltage step[24]. Indeed, quantum conductance corresponding to a dual-ion bridge assisted charge transport process can even be realized, as demonstrated in Fig. 3a. $R_{LRS}^{-1}$ manifests as *N*-fold single bridge conductance $G_s$, with the integer *N* reflecting the number of conductive dual-ion bridges.

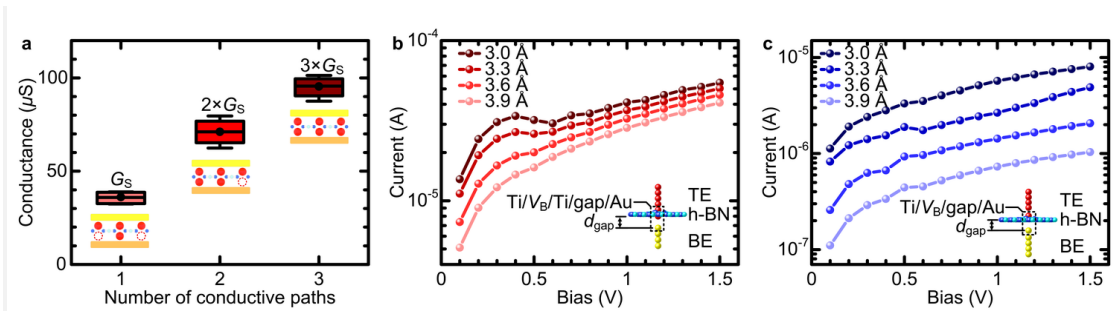

**Fig. 3: Sub-quantum conductance and discrepancy analysis in single conductive bridge. a**, Discrete LRS conductance levels in one device set by small sweep steps (~20 μV/step). Each level approximately equals an integer multiple of the single bridge conductance. This reflects the varying number of conductive bridges. Error bars indicate the 10-90% distribution, with points marking the arithmetic mean and horizontal lines representing the median. **b,c**, The first-principles calculations of transport properties with single conductive bridge in LRS (**b**) and HRS (**c**). $R_{LRS}$ and $R_{HRS}$ increase as the gap distance ($d_{gap}$) expand. This yields sub-quantum conductance as well as cycle-to-cycle and device-to-device variations.

However, the conductance value measured for a single dual-ion bridge is smaller than the theoretical quantum conductance ($G_0 = 2q^2/h$ ~77 μS) [25]. It is associated with the non-ideal contact (the gap as in Fig. 2b) at the vdW interface between the h-BN and BE. The scattering encountered as electrons traverse the interface gap attenuates the transmitted current[26]. The dependence of distinct conductance values on different gap distances ($d_{gap}$) is corroborated by the first-principles calculations shown in Fig. 3b. The sub-quantum conductance indicates the possibility of achieving lower energy consumption.

Fig. 3c exhibits the first-principles calculations of transport properties in HRS. The larger resistance, when the Ti ion only remains at the Base site without forming any conductive bridge, is consistent with the experimental $R_{HRS}$. Variations in $d_{gap}$ also contribute to the differences in $R_{HRS}$. Direct CVD grown h-BN on electrodes in future research is expected to eliminate the gap[27], thereby alleviating the variations.

**Ultra-fast switching and ultra-low energy consumption**

To enable the single-ion operation and verify the ultra-fast switching speed, the test setup shown in Fig. 4a is constructed. The arbitrary waveform generator is capable of producing ultra-short pulses with low noise amplifier (LNA) and radio frequency switch (RFSW). A 59-GHz high-frequency oscilloscope was used to monitor the ultra-short voltage pulses applied to the single-ion memory device (see Methods and Supplementary Information for experimental details). During the SET process, the applied pulse waveform with width $t_{SET}$ ~20 ps is presented in Fig. 4b. The single-ion memory switches on nonvolatilely, in agreement with the first-principles predictions in Fig. 1b. It was rather interesting to observe the first experimental conductance value ever measured for a single dual-ion bridge ($R_{LRS}^{-1}$ ~ 20 μS, Fig. 4d). It should be noted that the pulse width of 20 ps was limited by the measurement setup and equipment that are currently available, and the single-ion memory is supposed to switch within 10 ps, as the first-principles calculations suggest (Fig. 1b). In contrast, the RESET process is slower ($t_{RESET}$ ~200 ps, Fig. 4c), confirming that it involves Joule heating. Additionally, a relatively longer 130 ps SET voltage pulse forms more conductive dual-ion bridges, enlarging the ON/OFF ratio (Supplementary Fig. 14). Correspondingly, greater Joule heating is required to rupture these multiple conductive paths.

The 20 ps SET process in Fig. 4b yields an ultimate switching energy consumption of only ~310 aJ/bit, calculated using

$$E = \int \frac{V^2}{R} dt \quad (2)$$

This record-low value stems from the SIT path with low energy barrier, picosecond speed, and sub-quantum conductance. Notably, time-resolved resistance measurement is precluded by instrument resolution. The energy estimate is based on the minimum resistance value $R_{LRS}$. Therefore, the actual energy consumption is even lower.

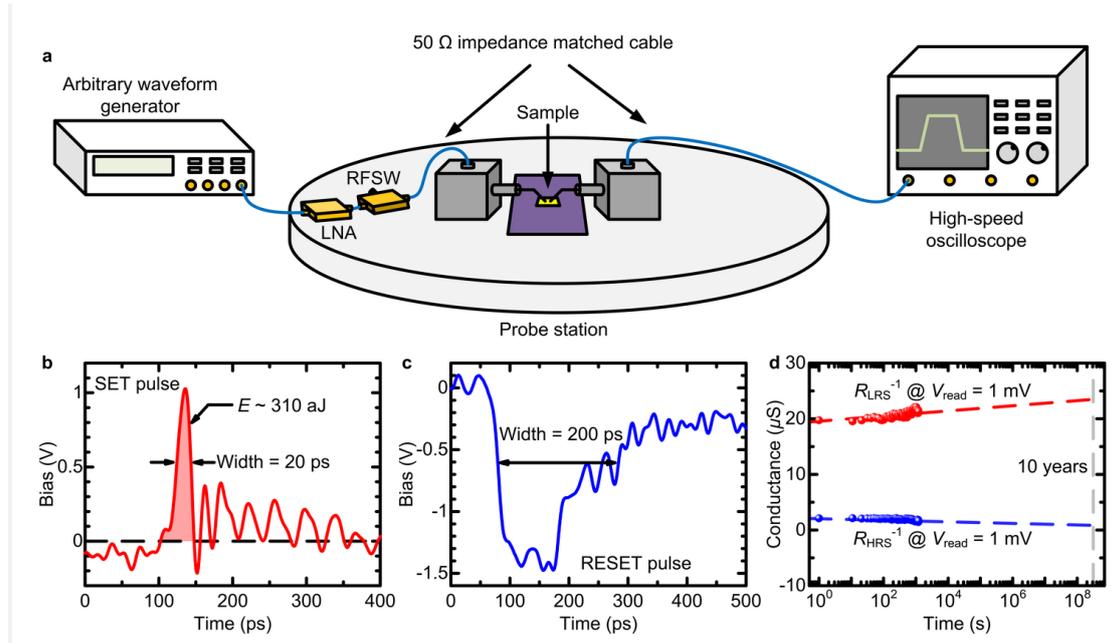

**Fig. 4: Ultra-fast switching experimental results. a**, The schematic of the high-frequency pulse test setup. The 50 Ω connecting cables should be kept as short as possible. **b,c**, The applied SET (**b**) and RESET (**c**) pulse waveforms measured when cascaded with THROUGH samples. The 3-dB pulse widths are 20 ps and 200 ps, respectively. **d**, Nonvolatile memory states in HRS and LRS after applying pulses. The sub-quantum conductance can be attributed to the ultra-low applied energy.

Compared with other nonvolatile memristors[11,14,15,28-45], including multilayer h-BN devices, the SIT mechanism in monolayer h-BN achieves unprecedented switching speed and ultra-low energy consumption (Fig. 5a and Supplementary Table 1). It even consumes less energy and operates several orders of magnitude faster than biological counterparts as in human brains ($E \sim 1$ fJ/spike and $t \sim 100$ ms/event)[45,46]. Among memory technologies, the proposed single-ion memory delivers the fastest switching speed for nonvolatile storage (Fig. 5b). The ultra-small cell size and vertical device structure make it intrinsically suited for high storage density and multiple layers of 3D stacking. The CVD based simple fabrication process is also compatible with standard complementary metal-oxide-semiconductor (CMOS), which makes it possible to integrate single-ion memory devices directly on top of computing circuits, and resolve the "Memory Wall" problem in an elegant way. Therefore, we believe we found the ideal solution for the "Unified Memory" as long desired by the industry.

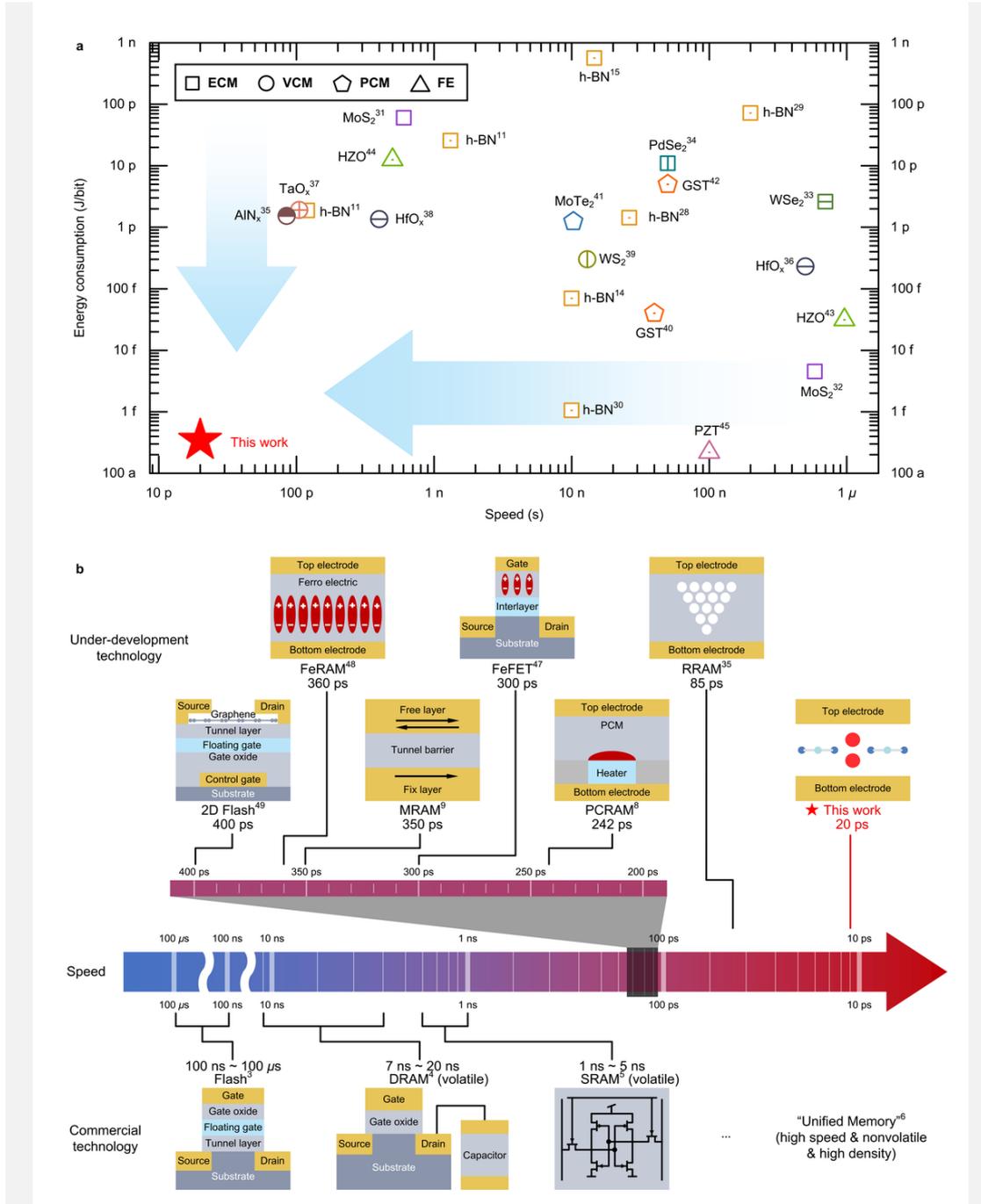

**Fig. 5: Summary of reported memories with different material and mechanism. a**, Monolayer h-BN based single-ion memory in this work achieves picosecond-scale programming speed and attojoule-per-bit switching energy consumption simultaneously. Compared with Electrochemical Metallization (ECM)[11,14,15,28-34], Valence Change Mechanism (VCM)[35-39], Phase Change Mechanism (PCM)[40-42], and Ferroelectric (FE) devices[43-45], this SIT mechanism exhibits the fastest speed and ultra-low energy, which is attributed to ultra-short migration distance and sub-quantum conductance. **b**, Among the commercial[3-5] and under-development technologies[8,9,47-49], single-ion memory offers the fastest switching speed for nonvolatile storage.

**Conclusion**

When the h-BN switching layer is reduced to a monolayer, first-principles calculations and experimental results reveal that the single-ion transport (SIT) mechanism would happen and dominate the switching process. Driven by the electric field and Joule heating, a single Ti ion penetrates the boron vacancy defect into the Trap site, while the vacated Base site is simultaneously occupied by another ion to establish a dual-ion conductive bridge. By virtue of the sub-nanometer migration distance and sub-quantum conductance, the memory device concurrently achieves 20 ps switching and sub-femtojoule energy consumption. This is the first demonstration of an electronic device that is based on the transport and trapping of a single ion at room temperature. This storage paradigm, relying on the manipulation of a single ion, offers a route to high-speed, low-energy, and high-density nonvolatile memory. The single-ion memory technology opens an avenue toward ultra-efficient and ultra-fast storage, computing, sensing, and intelligent systems in future.

## Methods

### First-principles calculations

Our first-principles calculations were performed by using the Atomic-orbital Based Ab-initio Computation at USTC (ABACUS) code. The Perdew-Burke-Ernzerhof (PBE) exchange-correlation functional was adopted, and the density functional theory with dispersion correction DFT-D3 was used to account for the van der Waals (vdW) interactions. The ABACUS code is developed to perform large-scale density functional theory calculations based on numerical atomic orbitals (NAOs). The optimized norm-conserving Vanderbilt (ONCV) fully relativistic pseudopotentials from the PSEUDODOJO library were used, and the double-ζ plus polarization functions (DZP) with a plane-wave cutoff energy of 100 Ry was employed as the NAO basis set. The valence electron configurations for Ti, Au, B and N were defined as $3s^23p^63d^24s^2$, $5d^{10}6s^1$, $2s^22p^1$, and $2s^22p^3$, respectively. The NAO bases for Ti, Au, B and N were set to $4s^2p^2d^1f$, $4s^2p^2d^1f$, $2s^2p^1d$, and $2s^2p^1d$, respectively.

The 8×8×1 supercell of monolayer h-BN with and without $V_B$ and $V_N$ with a 20 Å vacuum layer was constructed to study the defect-and-adsorption-related structures and properties. The energies and forces were computed with a kspacing of 0.1 Bohr$^{-1}$ and an electronic density convergence threshold of $1.0 \times 10^{-7}$ Ry. In this work, all atom positions of the initial state and the final state were fully optimized until all forces are less than 0.01 eV/Å. The nudged elastic band (NEB) method was used to calculate diffusion barrier energy and adsorption energy for the adsorption and migration of Ti ions. Nine interpolated intermediate images along the reaction path were created for all NEB calculations (force criterion is 0.1 eV/Å) to find the transition state and the barrier energy.

In addition, ab initio molecular dynamics (AIMD) simulations were carried out under an external electric field of 0.8 V/Å along the -z direction. All AIMD simulations were performed in the NVT ensemble at 400 K using a time step of 2 fs, with a total simulation time of 10 ps.

### Fabrication of the h-BN based devices

As shown in the schematic of Supplementary Fig. 7, the h-BN based devices were fabricated on sapphire substrates. First of all, photoresist was spin-coated on the cleaned substrate and the bottom electrodes with GSG structure were patterned by lithography. Then, 5 nm Cr and 50 nm Au were deposited by an electron beam evaporation system with 0.5 Å/s and 1 Å/s, respectively.

After lifting off, h-BN was transferred onto bottom electrodes by wet approach. Commercial monolayer h-BN used were grown on copper foil by CVD. PMMA was spin-coated on the copper foil as the supporting layer. After etching the copper foil with copper etchant, washed the PMMA/h-BN film at least three times with DI water to remove residual etchant. Next, the film floating on DI water was scooped up by substrate and naturally dried for 12 h. In the end, after baking on a 130 °C hotplate for 30 min, the PMMA was removed by acetone.

With photoresist as the mask, h-BN was etched 30 s by RIE in mixture of 40 sccm

CHF$_3$ and 4 sccm O$_2$ plasma with 60 W power, 50 mTorr pressure. Finally, after striping photoresist, top electrodes were patterned by lithography and deposited by an electron beam evaporation system with 20 nm Ti (0.5 Å/s) and 60 nm Au (1 Å/s).

**Physical characterization methods**

Atomic-resolution STEM and EDS were measured in FEI Themis Z. Raman spectroscopy with a 532 nm excitation source characterized the Raman spectra of h-BN at room temperature. Bruker Dimension Icon Atomic Force Microscope measured the thickness of h-BN with tapping mode. SEM were measured in Hitachi SU8220 with 3 kV accelerating voltage.

**Electrical measurements**

The *I-V* and retention curves of the h-BN based devices were measured on a Cascade Summit 200 probe station with a Keysight B1500A semiconductor parameter analyzer under ambient conditions. A heating platform was used for the variable temperature experiment.

A Keysight M8199B arbitrary waveform generator, a Keysight DSOZ594A high-speed oscilloscope, and customized LNA and RFSW were used for ultrafast switching measurement. The setup is linked via 50 Ω radio cables kept as short as possible to minimize insertion loss.

**Data availability**

All relevant data for this study are available in the article, the Source Data, and the Supplementary Information, and are available from the corresponding author upon reasonable request.

**Acknowledgements**

This work was supported by the National Natural Science Foundation of China (Grant Nos. 62231023 and 12134012), China Postdoctoral Science Foundation (Grant No. 2025M770515), the National Key Research and Development Program of China (Grant Nos. 2025YFE0201100 and 2023YFE0202300), the University Synergy Innovation Program of Anhui Province (Grant No. GXXT-2023-002), the Scientific Research Plan Program of Anhui Province (Grant No. 2024AH052041), and the University of Science and Technology of China (USTC) Research Funds of the Double First-Class Initiative (Grant No. YD2100002014). Preparation of h-BN was supported by Shanghai PrMat Technology Co., Ltd. This work was partially performed at USTC Center for Micro and Nanoscale Research and Fabrication, and was partially performed at USTC Integrated Circuit Laboratory, and was partially performed at USTC Instruments Center for Physical Science, and was partially performed at USTC National Synchrotron Radiation Laboratory. The numerical calculations were performed on Hefei advanced computing center and USTC HPC facilities.

## Author information

These authors contributed equally: Hengxiao Cheng, Xudong Zhu, Zijia Su, Zhongbin Dai, Jie Yu, Zhi Yan

Authors and Affiliations

**School of Integrated Circuits, University of Science and Technology of China, Hefei, China**

Hengxiao Cheng, Zijia Su, Zhongbin Dai, Jie Yu, Renfa Zhou, Juan Wang, Yuanyuan Shi, Zhongguang Xu, Chengjie Zuo

**Institute of Artificial Intelligence, Hefei Comprehensive National Science Center, Hefei, China**

**CAS Key Laboratory of Quantum Information, University of Science and Technology of China, Hefei, China**

Xudong Zhu, Lixin He

**School of Integrated Circuit, Hefei University of Technology, Hefei, China**

Zijia Su

**School of Chemistry and Materials Science & Key Laboratory of Magnetic Molecules and Magnetic Information Materials of Ministry of Education, Shanxi Normal University, Taiyuan, China**

Zhi Yan, Xujin Zhang

Contributions
H.C., Z.S., Z.D. and C.Z. conceived the research and designed the experiments. H.C. and R.Z. fabricated the devices and performed the measurements. X.Z., Z.Y., X.Z. and L.H. performed the first-principles calculations. H.C., X.Z., J.Y., Z.X. and C.Z. performed experimental data analysis and theoretical analysis. All authors contributed to the scientific discussions as well as the writing and revisions of the paper.

Corresponding authors
Correspondence to Zhongguang Xu, Lixin He or Chengjie Zuo.

## Competing interests

The authors declare no competing interests.

## Supplementary information

This file contains Supplementary Figs. 1-14, Supplementary Table 1, captions for Supplementary Videos 1-7, and Supplementary References 1-22.

# Supplementary Information

# Nonvolatile single-ion memory with picosecond switching

Hengxiao Cheng[1], Xudong Zhu[2,3], Zijia Su[1,4], Zhongbin Dai[1], Jie Yu[1], Zhi Yan[5], Xujin Zhang[5], Renfa Zhou[1], Juan Wang[1], Yuanyuan Shi[1], Zhongguang Xu[1]*, Lixin He[2,3]*, Chengjie Zuo[1]*

[1]School of Integrated Circuits, University of Science and Technology of China, Hefei, China
[2]Institute of Artificial Intelligence, Hefei Comprehensive National Science Center, Hefei, China
[3]CAS Key Laboratory of Quantum Information, University of Science and Technology of China, Hefei, China
[4]School of Integrated Circuit, Hefei University of Technology, Hefei, China
[5]School of Chemistry and Materials Science & Key Laboratory of Magnetic Molecules and Magnetic Information Materials of Ministry of Education, Shanxi Normal University, Taiyuan, China

*e-mail: xuxu@ustc.edu.cn, helx@ustc.edu.cn, czuo@ustc.edu.cn

## First-principles calculations

To explore the mechanism of monolayer h-BN based resistive switching, we performed the following first-principles calculations:

Initially, the migration of Ti ions on perfect h-BN surface was investigated. The lower diffusing energy barriers (~0.3 eV for $Ti^{3+}$, Supplementary Fig. 1 and Supplementary Video 1) indicate that Ti ions can easily diffuse on the h-BN surface.

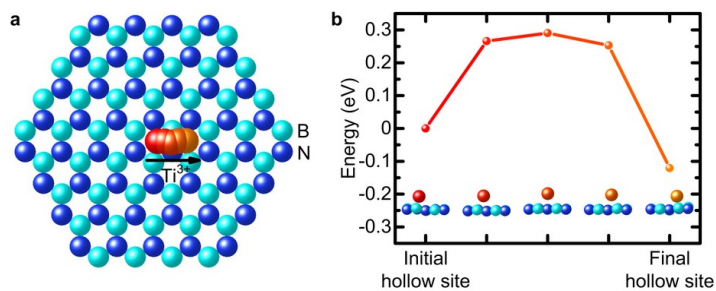

**Supplementary Fig. 1 | The simulations of $Ti^{3+}$ migrating on defect-free h-BN surface. a**, Ti ion migrates from one hollow site to another along the lowest energy pathway. **b**, The energy variation at different positions relative to the hollow site is slight, indicating that diffusion on defect-free h-BN surface is remarkably facile. The calculated dynamic distribution is shown in Supplementary Video 1.

Subsequently, we analyzed the energy associated with the migration of Ti ion from a hollow site to a relaxed site within the adjacent defect. As depicted in Supplementary Fig. 2 for $V_B$ defect, it is evident that once ion overcomes a relatively small barrier, there is a precipitous decline in energy, signifying trapped by the defect site (from hollow site to top relaxed site). To escape the defect (from top relaxed site to hollow site), a significantly larger energy is required. Ultimately, the energy minimum where Ti ion stabilize (defined as relaxed site) is located on either side of the BN plane. Meanwhile, a lower energy required to penetrate through the BN plane (from top relaxed site to bottom relaxed site).

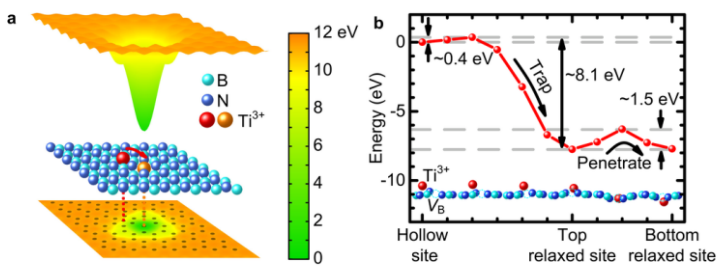

**Supplementary Fig. 2 | The energy variation of $Ti^{3+}$ migration for $V_B$ defect. a**, The adsorption energy landscape about Ti ion's migration on h-BN with a $V_B$ defect. **b**, The lowest energy pathway of $Ti^{3+}$ migration from hollow site to bottom relaxed site. The insets show the cross-sectional distributions obtained from simulations. Ti ion is easily trapped at the minimum energy sites (defined as relaxed site) located on either side of the defect. Penetrating the $V_B$ defects requires overcoming an energy barrier. The calculated dynamic distribution is shown in Supplementary Video 2.

In contrast, the energy barriers for $Ti^{3+}$ entering and leaving $V_N$ defect are small, and the former is even higher than the latter (Supplementary Fig. 3). It indicates that the Ti ion is not easily trapped by $V_N$ defects. This may be related to the positively charged dangling bonds of adjacent boron atoms. The repulsive force also increases the energy required for Ti ion when migrating from relaxed site to in-plane site within $V_N$ defect (~5 eV, Supplementary Fig. 4). As shown in the inset in Supplementary Fig. 4, Ti ion needs to repel surrounding boron atoms when passing through in-plane site. It indicates that Ti ion migration at $V_B$, rather than $V_N$, leads the resistive switching phenomenon.

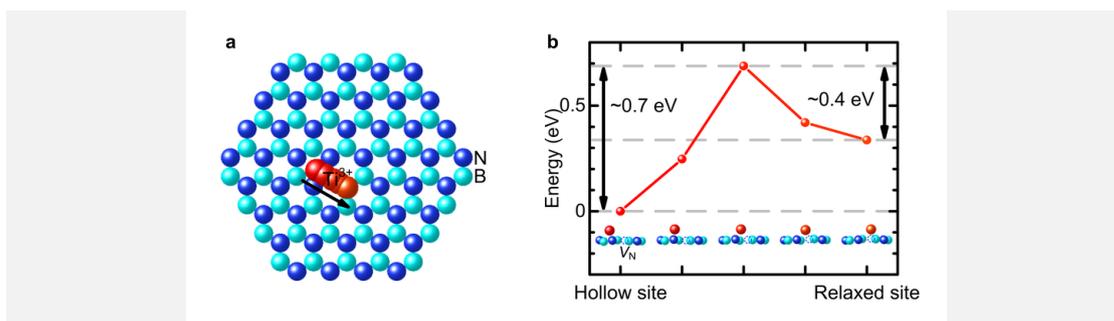

**Supplementary Fig. 3 | The lowest energy pathway of $Ti^{3+}$ migrating from hollow site to relaxed site within $V_N$ defect. a**, Ti ion migrates from hollow sites in adjacent hexagonal ring to relaxed sites along the lowest energy pathways on h-BN surface with $V_N$ defect. **b**, The energy barrier of $Ti^{3+}$ entry into the relaxed site in $V_N$ defect is larger than escape the site. The calculated dynamic distribution is shown in Supplementary Video 3.

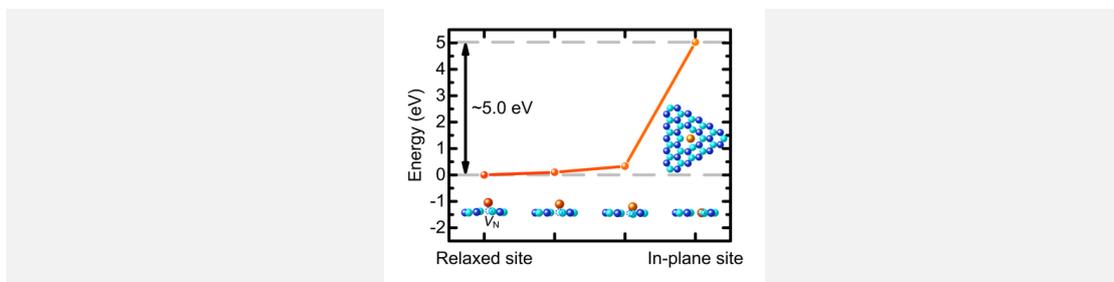

**Supplementary Fig. 4 | The energy barrier of $Ti^{3+}$ migrating from relaxed site to in-plane site within $V_N$ defect.** Ti ion penetrating vertically through $V_N$ defect needs large energy. The insets show the distributions obtained from simulations. The calculated dynamic distribution is shown in Supplementary Video 4.

We also calculated the complete dual-ion case, involves a Ti ion A (red ball in the inset) traversing $V_B$ defect to the bottom relaxed site (defined as the Trap site) with the vacated top relaxed site (defined as the Base site) replenished by another ion B (pale red ball in the inset). Due to the occupation of the initial Trap site by ion A, the position of ion B remains uncertain. When Ti ion B is initially located at the No. 1 hollow site in Supplementary Fig. 5a, the SET energy barrier is about 1.9 eV (Fig. 1b in manuscript). However, when ion B is initially located at No. 2 hollow site in Supplementary Fig. 5a, the SET energy barrier is about 1.0 eV (Supplementary Fig. 5b). The penetration barrier varies depending on the initial distribution of Ti ions. The reverse barrier from the final site back to the initial site is larger than that from the initial site to the final site. It confirms that the RESET process demands more energy and is assisted by Joule heating.

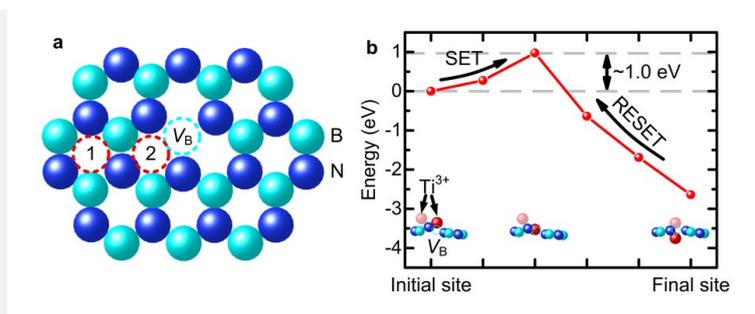

**Supplementary Fig. 5 | The lowest energy pathway of complete dual-ion migration. a,** The distribution diagram with $V_B$ on monolayer h-BN. In the SET process, a Ti ion A traverses $V_B$ defect to bottom relaxed site and another ion B occupies the top relaxed site. The initial ion B position in Fig. 1b in manuscript is No. 1 hollow site. **b,** The different energy barrier when the ion B is initially located at No. 2 hollow site. The insets show the distributions obtained from simulations. The calculated dynamic distribution is shown in Supplementary Video 6.

In order to further determine that metal ion crossing h-BN is responsible for resistive switching, we analyzed transport properties of LRS and HRS. When a Ti ion cross the BN plane, Ti ions present on both sides of the h-BN sheet (Fig. 3b in manuscript, Ti/h-BN/Ti/gap/Au). The conductive path generates a small conductance on the order of $10^{-4}$ S, which corresponds to the quantum conductance level ($G_0$ ~77 $\mu$S) and decreases with increasing $d_{\text{gap}}$. When Ti ion has not passed through the defect (Fig. 3c in manuscript, Ti/h-BN/gap/Au), the lower current corresponds to the HRS state current. $R_{\text{HRS}}$ increases monotonically with the $d_{\text{gap}}$. These correlations provide plausible explanations for the observed sub-quantum conductance and device-to-device variations.

Finally, the time-resolved complete migration process is simulated dynamically as shown in Fig. 1b in manuscript and the Supplementary Video 7. At the 8th picosecond, one Ti ion penetrate across the BN plane and being trapped by the Trap site, while another one occupies the vacant Base site. This provides theoretical support for ultrafast switching.

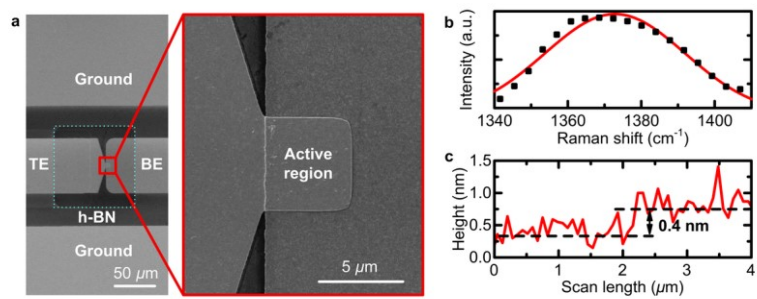

**Supplementary Fig. 6 | Characterizations of monolayer h-BN based devices. a**, Scanning electron microscopy (SEM) images of CVD h-BN devices with 5 $\mu$m × 5 $\mu$m active region. **b**, Raman spectrum of CVD-grown monolayer h-BN. The red solid line represents the Gaussian fit to experimental data. **c**, AFM morphology image at the edge of CVD h-BN film.

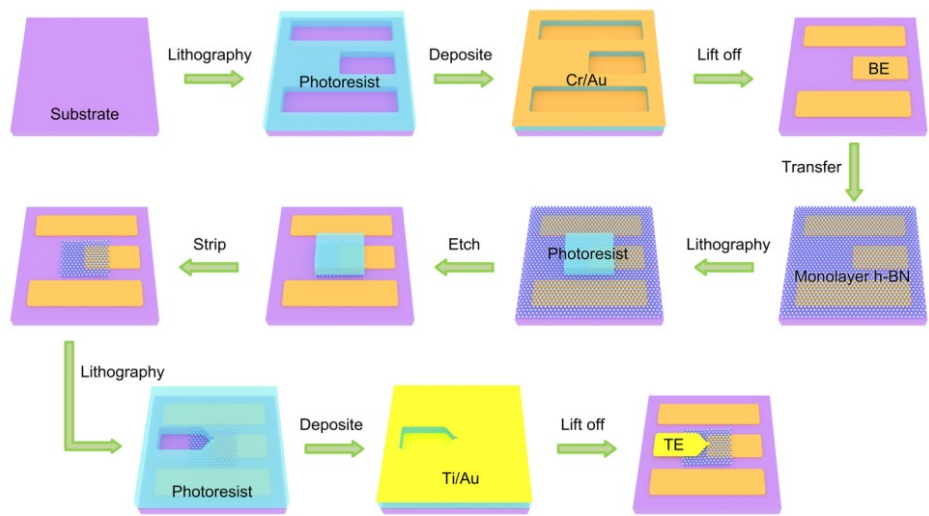

**Supplementary Fig. 7 | The fabrication schematic illustrations of the h-BN based devices.** The bottom electrodes are fabricated by lithography, deposition, and lifting off, followed by the wet transfer process and etching of h-BN. The manufacturing of the top electrodes is similar to the bottom electrodes. It should be noted that the dimensions presented in the schematics are intentionally magnified for illustrative purposes and don't correspond to the actual dimensions of the devices.

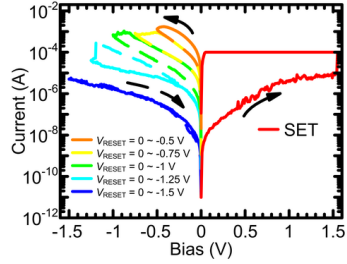

**Supplementary Fig. 8 | The *I-V* curves with different RESET sweep waveforms after a SET process.** The quantity of ruptured conductive paths is affected by the $V_{RESET}$ range. Take advantage of the phenomenon, multilevel storage can be implemented.

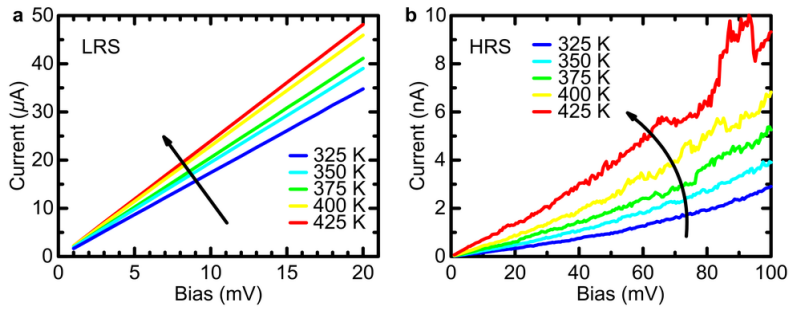

**Supplementary Fig. 9 | The *I-V* curves in CVD monolayer h-BN based device with different temperatures.** The results in LRS (**a**) and HRS (**b**) reveal semiconductor-like behavior, with the resistances exhibiting negative temperature coefficients that decrease as temperature increase.

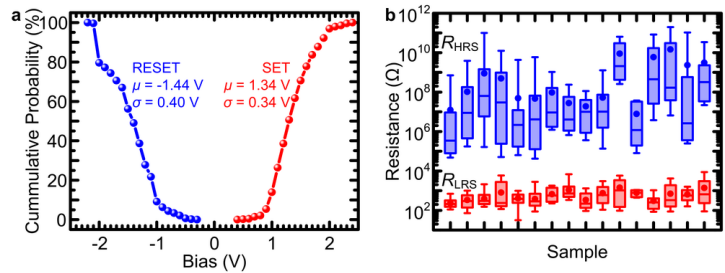

**Supplementary Fig. 10 | Statistical analysis in the h-BN based resistive switching.** **a**, The cumulative probability plot of $V_{SET}$ and $V_{RESET}$ extracted from 90 devices across 9 batches. **b**, The resistance statistics of $R_{HRS}$ and $R_{LRS}$ from 16 samples. Error bars indicate the 10-90% distribution, with points marking the arithmetic mean and horizontal lines representing the median.

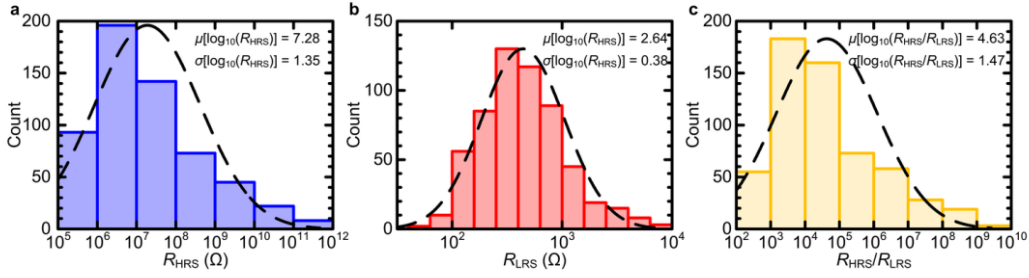

**Supplementary Fig. 11 | The distributions of $R_{HRS}$, $R_{LRS}$, and $R_{HRS}/R_{LRS}$ from 90 monolayer h-BN BRS samples across 9 batches.** The dashed Lines represent the normal fits to the experimental data on logarithmic scale. The mean values for $R_{HRS}$ (**a**), $R_{LRS}$ (**b**), and $R_{HRS}/R_{LRS}$ (**c**) are about $3.80\times10^7$ Ω, $4.37\times10^2$ Ω, and $4.26\times10^4$, respectively. The observed device-to-device and cycle-to-cycle variations are within acceptable limits.

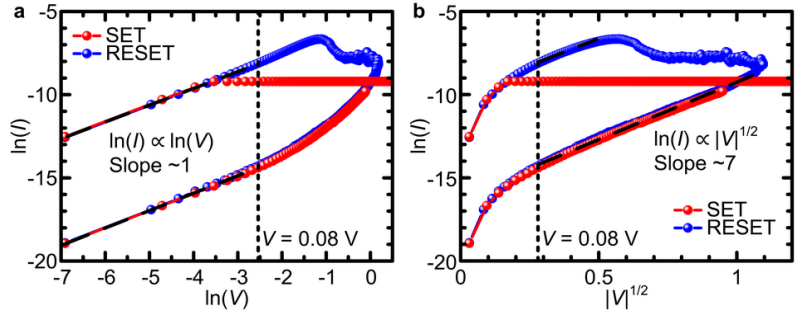

**Supplementary Fig. 12 | The ln($I$) vs. ln($V$) and ln($I$) vs. $V^{1/2}$ curves of monolayer h-BN based BRS. a**, In the low voltage range, the current through Au/Ti/monolayer h-BN/Au device is proportional to the voltage, as evidenced by the slope of approximately 1 for ln($I$) vs. ln($V$). This indicates that the dominant conduction mechanism is ohmic conduction. **b**, In the high voltage range, ln($I$) ∝ $V^{1/2}$ satisfies the Schottky equation [S1]:

$$\ln(I) = \left[\ln\left(\frac{4\pi q m^*(kT)^2}{h^3}\right) - \frac{q\phi_B}{kT} + \ln(S)\right] + \frac{q^{3/2}}{2kT\sqrt{\pi\varepsilon t}}V^{1/2} \quad (S1)$$

where $m^*$, $\phi_B$, $\varepsilon$, $S$, and $t$ denote the electron effective mass, Schottky barrier height, dielectric constant, active area, and thickness of the h-BN switching layer, respectively. $h$ is the Planck's constant. The slope of ln($I$) vs. $V^{1/2}$ is about 5 to 10 in the range above 80 mV. Energetic stimulation drives electron injection into the conduction band and generates current. In this high voltage range, the dominating conduction mechanism becomes Schottky emission.

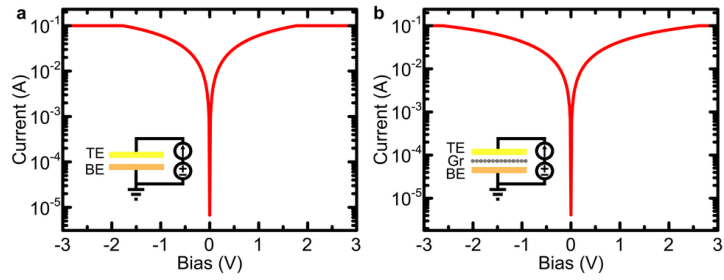

**Supplementary Fig. 13 | The *I-V* curve in reference samples. a,** The result of Au/Ti/Au stack without h-BN does not exhibit obvious memristive characteristics. Therefore, the effect of metal oxides can be neglected. **b**, The *I-V* curve in devices with Au/Ti/graphene/Au stack. CVD monolayer graphene grown on copper foil was wet-transferred using the same process as h-BN. The curve without memristive characteristic indicates that BRS originate from h-BN rather than impurities introduced during the transfer process.

**Ultra-short pulse test**

To measure the switching speed of the memory, we used the setup shown in Fig. 4a in manuscript to apply pulse and then detect changes in the resistive state. A pulse sequence with period of 1 $\mu$s and width of 20 ps was generated by an arbitrary waveform generator Keysight M8199B. The 20 ps pulse width represents the limit of the generator's output capability. The signal was transmitted through 50 $\Omega$ RF cables with an attenuation of 5 dB/m @ 50 GHz, necessitating the shortest possible cable length. The pulse sequence was then amplified by a low noise amplifier (LNA) with a frequency range of 50 kHz ~ 67 GHz, a gain of 25 dB, and a linear output power (output P1dB) of 15 dBm. Subsequently, a single-pole-double-throw RF switch (SPDT-RFSW) was used to extract one pulse for application to the sample. The switch operates at 100 kHz ~ 50 GHz with an insertion loss of 3 dB and a switching speed of 100 ns. The two output ports of the RFSW were connected to a 50 $\Omega$ matched terminator and to the sample via a probe, respectively. The final output waveform was read by a high-speed oscilloscope Keysight DSOZ594A. The bias waveform was read when the probe was contacted with the control sample without h-BN (regarded as THROUGH). When cascaded with the h-BN based devices in the same setup, the bias was almost entirely applied between the TE and BE due to the large resistance.

As shown in Supplementary Fig. 14, when the SET pulse width is broadened to 130 ps, the $R_{LRS}$ decreases to 220 $\Omega$ and the ON/OFF ratio increases to >10$^5$. This indicates that higher energy drives more ions migration, thereby creating additional conductive paths. Correspondingly, a wider RESET pulse (~800 ps) is required to switch off the memory.

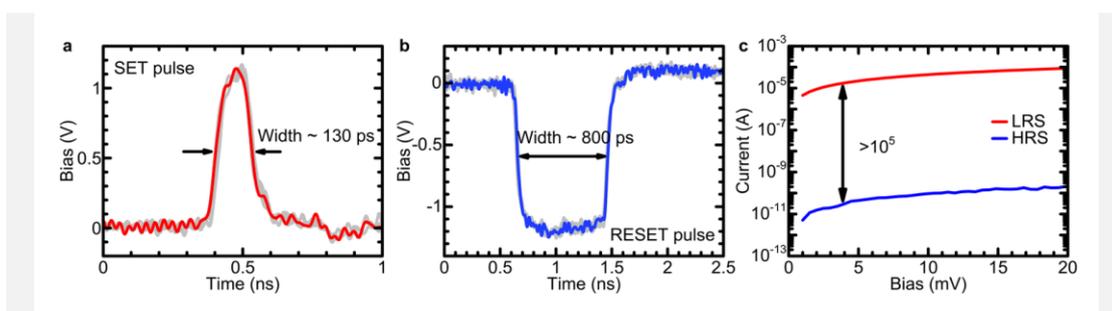

**Supplementary Fig. 14 | Experimental switching results with large ON/OFF ratio. a,b**, The SET (**a**) and RESET (**b**) pulse waveform with width of 130 ps and 800 ps, respectively. **c**, The *I-V* curves in LRS and HRS measured after applied pulses. The ON/OFF ratio of >10$^5$ and $R_{LRS}$ of 220 $\Omega$ indicate the formation of numerous conductive paths.

**Supplementary Table 1 | The comparisons of nonvolatile memristors.**

| References | Structure | RS mechanism | RS thickness | Switching Speed | Energy Consumption* |
|---|---|---|---|---|---|
| S2 | Ti/h-BN/Au | ECM | 2.5 nm | 1.32 ns | 25.1 pJ/bit |
|  |  |  | 1.2 nm | 120 ps | 1.88 pJ/bit |
| S3 | Ti/h-BN/Au | ECM | 5 nm | 30 ns | 1.2 pJ/bit |
| S4 | Ag/h-BN/Au | ECM | 6 nm | 200 ns | 72 pJ/bit |
| S5 | Ti/h-BN/Gr | ECM | 5 nm | 10 ns | 70 fJ/bit |
| S6 | Ag/h-BN/p$^{++}$Si | ECM | 2.7 nm | 10 ns | 1.05 fJ/bit |
| S7 | Cr/h-BN/Au | ECM | 0.4 nm | 15 ns | 680 pJ/bit |
| S8 | Cr/MoS$_2$/Au | ECM | 0.65 nm | 500 ps | 49 pJ/bit |
| S9 | Ag/MoS$_2$/Ag | ECM | 50 nm | 600 ns | 4.5 fJ/bit |
| S10 | Ag/WSe$_2$/Ag | ECM | 2.2 nm | 700 ns | 2.6 pJ/bit |
| S11 | Ti/PdSe$_2$/Au | ECM | 3 nm | 50 ns | 11 pJ/bit |
| S12 | TiN/AlN$_x$/Pt | VCM | 6 nm | 85 ps | 1.53 pJ/bit |
| S13 | TiN/HfO$_x$/Gr | VCM | 5 nm | 500 ns | 230 fJ/bit |
| S14 | Pt/TaO$_x$/Pt | VCM | 7nm | 105 ps | 1.9 pJ/bit |
| S15 | TiN/HfO$_x$/TiN | VCM | 8 nm | 400 ps | 3.6 pJ/bit |
| S16 | Pt/WS$_2$/Pt | VCM | 2 nm | 13 ns | 299 fJ/bit |
| S17 | Pt/GST/Pt | PCM | 100 nm | 40 ns | 43 fJ/bit |
| S18 | Ti/MoTe$_2$/Au | PCM | 6 nm | 10 ns | 1 pJ/bit |
| S19 | M/GST/M | PCM | 100 nm | 50 ns | 5 pJ/bit |
| S20 | Ti/HZO/W | FE | 6 nm | 1 μs | 31.25 fJ/bit |
| S21 | Pt/SiO$_2$/HZO/TiN | FE | 2.5 nm | 500 ps | 12.5 pJ/bit |
| S22 | Ag/PZT/LSMO/STO | FE | 4 nm | 100 ns | 220 aJ/bit |
| This work | Ti/h-BN/Au | SIT | 0.4 nm | 20 ps | 310 aJ/bit |

*Based on the presented experimental results, the energy consumptions are calculated by uniformly using $E = \int \frac{V^2}{R} dt$.

**Supplementary Video 1 | Single Ti$^{3+}$ migrates from one hollow site to another on defect-free h-BN surface.** It corresponds to Supplementary Fig. 1.

**Supplementary Video 2 | Single Ti$^{3+}$ migrates from hollow site to Trap site on h-BN with $V_B$ defect.** It corresponds to Supplementary Fig. 2.

**Supplementary Video 3 | Single Ti$^{3+}$ migrates from hollow site to relaxed site on h-BN with $V_N$ defect.** It corresponds to Supplementary Fig. 3.

**Supplementary Video 4 | Single Ti$^{3+}$ migrates from relaxed site to in-plane site on h-BN with $V_N$ defect.** It corresponds to Supplementary Fig. 4.

**Supplementary Video 5 | The complete SIT switching on h-BN with $V_B$ defect.** It corresponds to Fig. 1b (top) in manuscript.

**Supplementary Video 6 | The complete SIT switching on h-BN with $V_B$ defect.** It corresponds to Supplementary Fig. 5.

**Supplementary Video 7 | The time-resolved SIT switching on h-BN with $V_B$ defect.** It corresponds to Fig. 1b (bottom) in manuscript.

# Supplementary References